# Travelling-wave, Quasi-periodic, and Longulent States of the Galerkin-regularized Hydrodynamic-type Systems

Jian-Zhou Zhu (朱建州)*

Su-Cheng Centre for Fundamental and Interdisciplinary Sciences, Gaochun, Nanjing 211316, China

# Abstract

Travelling-wave, quasi-periodic and "longulent" states of the Galerkin-regularized systems preserving finite Fourier modes are exposed. The longulent states are characterized by solitonic structures, called "longons", accompanied by disordered components, which is associated to whiskered tori according to the a-posteriori Kolmogorov-Arnold-Moser (KAM) theorem. On-torus invariants are introduced for constructing the KAM tori, towards a potential pseudo-integrability theory. Persistence of the longulent states with respect to certain dispersive and dissipative perturbations are also numerically indicated.

The Galerkin truncation/regularization (Gr) preserving finite Fourier modes regularizes the otherwise rough/singular compactons, peakons [1–4] and the more familiar Burgers-Hopf (BH) shocks, resembling yet differing from the linearly dispersive Korteweg-de Vries (KdV) regularizaiton. Such or similar truncations are widely used in analysis, computation, and effective field theories.

Let $v(x,t)$ solve, with $x$-period $2\pi$ and $v_0 = v(x,0)$,

$$v_t + vv_x = a. \tag{1}$$

As models in plasma and hydrodynamic nonlinear waves, and quantum shocks, among others, $a = 0$, $\mp v_{txx}/9 \mp 2v_x v_{xx}/27 \mp vv_{xxx}/27$, $\mu v_{xxx}$, and $v(v^2)_{xxx}$ identify, respectively, the BH, compacton/peakon (CP [1, 2]), KdV and Rosenau-Hyman [4] compacton [RH or $K(2,2)$] equations. In the CP model [reverting to the convenient form $v_t \pm v_{txx} + 3vv_x = \mp 2v_x v_{xx} \mp vv_{xxx}$ with the $x$-rescaling factor 3 from now on], the upper signs correspond to the compacton, and the lower to the Camassa-Holm (CH) peakon model [3].

Let us consider the problem in the period $[0, 2\pi)$. We have $\hat{v}_k = \int_0^{2\pi} \frac{v}{2\pi} e^{-\hat{i}kx} dx$, with complex conjugacy ($c.c.$) $\hat{v}_k^* = \hat{v}_{-k}$ for reality. Additionally, $v\partial_x v = \sum_k \hat{b}_k e^{\hat{i}kx}$ where $\hat{b}_k = \frac{\hat{i}k}{2} \sum_p \hat{v}_p \hat{v}_{k-p}$ and $\hat{i}^2 = -1$. For $v_0$ well-prepared in ${}^K\mathbb{G} = \{k : \ -K \le k \le K\}$ ("Galerkin space" hereafter), we can calculate each extra-Galerkin $\hat{b}_m$ for $K < |m|$ ($\le 2K$) from the intra-Galerkin $\hat{v}_k$s for $k \in {}^K\mathbb{G}$. In the BH case, setting $\hat{a}_m$ to be ${}^K\hat{g}_m = \hat{b}_m$ for $m \notin {}^K\mathbb{G}$ and 0 otherwise results in the Galerkin truncation: for all $m \notin {}^K\mathbb{G}$, $\hat{v}_m(t) \equiv 0$ ($t > 0$), thus the Galerkin-regularized BH (GrBH), and GrKdV with the same ${}^Kg$ form; and, similarly, the GrCP, or, GrCH and GrRH systems with their respective ${}^Kg$.

To be more explicit, we write down the equation for the above GrBH system, with

$P_K v(x) := \sum_{|k|\le K} \hat{v}_k \exp\{\hat{i}kx\} =: u$, $B := u^2/2$ and ${}^K G := B - P_K B$ [5],

$$Du/Dt := \partial_t u + \partial_x B = \partial_x {}^K G;\ u_0 = P_K v_0. \tag{2}$$

The Galerkin force ${}^K g = \partial_x {}^K G$, with ${}^K \hat{g}_m$ for $K < |m| \le 2K$, can be excited only when there exists $\hat{u}_k \neq 0$ with $k > K/2$. Such ${}^K g$ and other Gr-system ones are nonlinear dispersions.

The CP Hamiltonian operator $J_{CP} = -2\pi(\partial_x \pm \partial_x^3)$ in Fourier representation still applies with truncation and is inherited by GrCP, just like $J_{KdV} = -2\pi\partial_x$ by GrKdV [6]. The GrBH reduction replaces the Hamiltonian $\mathcal{H}_{BH} = \int_0^{2\pi} \frac{v^3 dx}{12\pi}$ with [7]

$$\mathcal{H} = \sum_{p,q,k=p+q\in {}^K\mathbb{G}} \frac{\hat{u}_k^* \hat{u}_p \hat{u}_q}{6} = \int_0^{2\pi} \frac{u^3 dx}{12\pi}. \tag{3}$$

${}^K\mathcal{G} = \mathcal{H}_{BH} - \mathcal{H}$ is called the Galerkin interaction potential. The alternative BH Hamiltonian operator $J'_{BH} := -(u\partial_x + \partial_x u)/3$ involves $u$ and is not transferable to GrBH. Only three GrBH invariants, $\mathcal{H}$, $\mathcal{E} = \int_0^{2\pi} \frac{u^2 dx}{4\pi}$ and $\mathcal{M} = \hat{u}_0 = \int_0^{2\pi} \frac{u dx}{2\pi}$ are known to survive for general $K$ ("rugged"); similarly for the GrKdV and GrCP situations, with, e.g., $\mathcal{H}_{CP} = \int_0^{2\pi} \frac{v^3 \mp v(\partial_x v)^2}{4\pi} dx$ and, accordingly, $\mathcal{M}_{CP}$ and $\mathcal{E}_{CP}$, and, their Gr-versions. By Galilean invariance, $\mathcal{M}$ is taken to be zero or truncated in this study, and $K$ is effectively the number of available modes (pairs of conjugate Fourier components) with $2K$ degrees of freedom.

Our main interests here are mainly the dynamical nature of the truncating force ${}^K g$ and the possibility of bridging integrability and nonintegrability of Hamiltonian systems. The objectives being the stable solitonic structures in physical space and the Kolmogorov-Arnold-Moser (KAM) tori in phase space, we numerically tested the travelling-wave solutions and found instabilities lead to states dominated by solitonic structures (that we call "longons") admist weaker less-ordered components, the "longulent states" corresponding to presumably whiskered tori. The mechanism for the associated de-thermalization appears akin to that of integrable systems, beyond the "Hamiltonian effect" concerned earlier [7, 8], which motivated proposing the notion of "pseudo-integrability" with appropriate on-torus or torus-specific but not rugged invariants, the latter being explicitly used for constructing multi-phase (quasi-)periodic solutions analytically. In particular, a-posteriori KAM theorem [9] is suggested, and the persistence of the KAM tori is demonstrated numerically. The scenario appears universal over all the hydrodynamic type models examined here.

Travelling waves and interacting longons — We begin with Gr-system example/special solutions of the travelling-wave form, $u^{\#}(x,t) = u^{\#}(\zeta)$ with $\zeta = x - \lambda t$, satisfying

$$\partial_t \hat{u}^{\#}_k = -\frac{\hat{i}k}{2} \sum_{p,q,p+q=k\in {}^K\mathbb{G}} \hat{u}^{\#}_p \hat{u}^{\#}_q = -\hat{i}\lambda k \hat{u}^{\#}_k. \tag{4}$$

For $\lambda = 0$, immediate examples include those with a single mode in $(K/2, K]$; while, with an arbitrary phase parameter $x_0$, $u^{\#} \propto 2\cos[K(x-x_0)/3] - \cos[K(x-x_0)]$ for $\mathrm{mod}(K,3) = 0$ are less-trivial ones more of which we will come back to later. With $x_0 = 0$ henceforth, we note that, for instance, $u^{\#} \propto 2\cos(2x) - \cos(6x)$ also solves Eq. (4) in ${}^7\mathbb{G}$, besides ${}^6\mathbb{G}$, without excitating/occupying $|k| = 7$; so, another wavenumber parameter $S$ for the maximal $|k|$ occupied by $u^{\#}$ is naturally introduced below. For moving waves ($\lambda \neq 0$) occupying $L$ modes equally spaced in $|k|$, with $\mathrm{mod}(S,L) = 0$ and $S \le K < (L+1)S/L$, we find, by straightforward calculation with $L = 2$,

$$u^{\#} = 2\sqrt{2}|\lambda|\cos(S\zeta/2) + 2\lambda\cos(S\zeta). \tag{5}$$

Others similarly follow. For instance, taking $\lambda > 0$, with $\theta = S\zeta$, we have a three-mode-occupation wave,

$$u^{\#} = \lambda[-2\chi_1\cos(\theta/3) + \chi_2\cos(2\theta/3) - \chi_1\chi_2\cos\theta] \tag{6}$$

where $\chi_1 = \sqrt{\frac{5-\sqrt{5}}{5}}$ and $\chi_2 = \sqrt{5} - 1$.

Accordingly, ${}^K g^{\#} \propto S\lambda^2$; for example, corresponding to Eq. (5),

$${}^K g^{\#} = -(S\lambda^2)[3\sqrt{2}\sin(3\theta/2) + 4\sin(2\theta)]/2. \tag{7}$$

Eq. (5) extends to GrKdV-GrRH waves, with both the $\nu$ and $\mu$ terms for unification,

$$u^{\#} = 2\sqrt{\frac{\lambda - \mu S^2}{1 + 2\nu S^2}\chi}\cos\frac{\theta}{2} + \chi\cos\theta \tag{8}$$

with $\chi = (4\lambda - \mu S^2)/(2 + \nu S^2)$; and, for the GrCP model,

$$\frac{u^{\#}}{\lambda} = \frac{4}{3}\sqrt{\frac{2(1 \mp S^2)}{(4 \mp S^2)}}\cos\frac{\theta}{2} + \frac{2}{3}\cos\theta. \tag{9}$$

Note that there is already the linearly dispersive regularization of BH in KdV which, however, unlike the odd-order-nonlinearity cases such as the cubically-nonlinear modified KdV (mKdV) or nonlinear Schrödinger (NLS) system, there is no single-mode or monochromatic (condensate) GrKdV solution that would be independent of the truncation threshold $K$; so, in the situation for the above $u^\#$ being valid, it does not make sense to talk about the convergence to KdV of GrKdV.

$u^\#$s occupying many $|k|$s may be obtained numerically or with the help of symbolic-computation softwares; e.g., with $\mathrm{mod}(S, 4) = 0$, approximate $\eta$s ($\{\eta_1, \eta_2, \eta_3, \eta_4\}$s) can be found in the ansatz

$$\frac{u^\#}{2\lambda} \approx \eta_1 \cos\frac{\theta}{4} + \eta_2 \cos\frac{\theta}{2} + \eta_3 \cos\frac{3\theta}{4} + \eta_4 \cos\theta. \tag{10}$$

Specifically, $\eta \approx \eta^c = \{-0.507, 0.450, -0.376, 0.292\}$ correspond to a GrBH case resembling the cnoidal wave, except for weaker wiggles between the strong pulses, which is also the case for Eqs. (8) and (9). [It is relatively easy to find by hand some very specific solutions for $S$ up to 7, and many more general ones can be found with the help of symbolic computation software quickly for such an $S$.]

Wiggle counts grow with mode numbers, as shown in Fig. 1's top and middle panels, using $K = S$ [always so below: other $K$s ($< S + S/L$) yield the same scenario, modulo quntitative variances.] Such (time-)periodic $u^\#$-orbits, together with other quasi-periodic tori to be constructed later, are associated to interacting solitonic structures, i.e., the "longons" to be explained eventually, akin to the $x$-periodic KdV (multi-)soliton analogue [11, 12].

Pseudo-spectral computations (discussed later) indicate that, as shown in Fig. 2, for $L \geq 2$ with unoccupied $k$s in the respective $u^\#_0$, instability eventually transforms the waves into robust states with interacting strong solitonic 'longons' amid weaker, less-ordered components of various propagating speeds roughly proportional to the signed strengths, featuring more widely separated crests and troughs whose levels vary slightly: such "longulent" states correspond presumably whiskered tori and, with respect to time, are "pseudo-periodic" in the sense that the weaker components introduce deviations from strict (quasi-)periodicity. The apparently solitonic longons locally resemble the Mexican hat. This is even clearer in larger-$K$ cases, as exemplified later in Fig. 3. There are waves of other shapes, with various organizations of the strong pulses and weak wiggles indicating some (yet unknown) symmetry in the solutions, which were observed in the developing phases but not in the developed/mature solitonic longons.

Mature longons exhibit a half-wavelength ${}^K g$ oscillation within each strong(est) pulse, similar to the case of travelling-wave $u^\#$s. This feature is universal across all Gr-systems studied here, with only minor differences in details such as strengths, as partially shown in

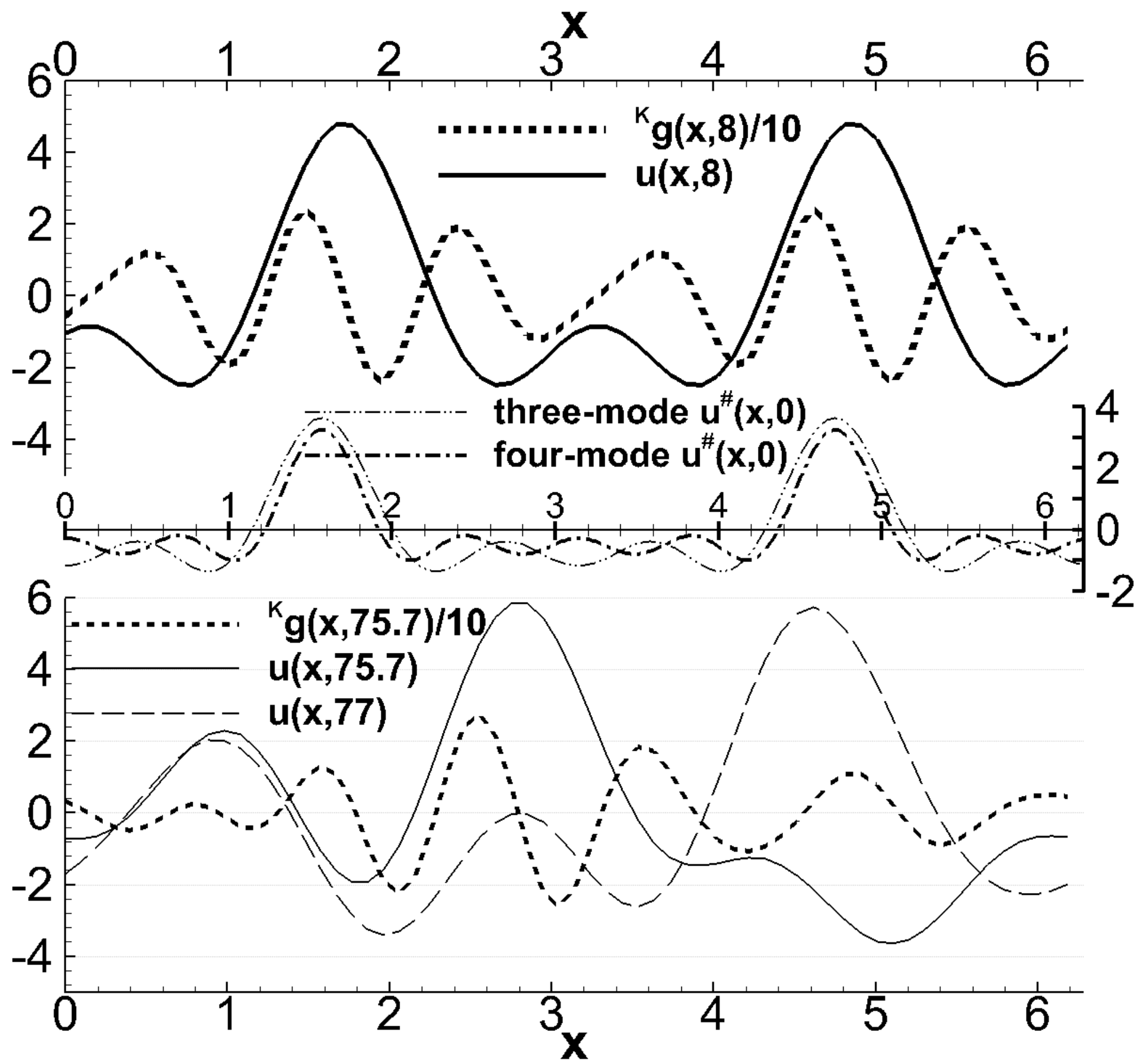

Figure 1. GrBH fields for $K = 4$ and $\lambda = 1$ of the two-active-mode solitary wave, at two regimes (upper and lower frames): $u(x,t)$ at $t = 8$ obeying Eqs. (5) transitioning to an interacting-longon state at $t = 75.7$ [with $u(x,77)$ added particularly to show varying crest and trough levels]. The middle frame for the three- and four-mode-occupation $u^\#$s with $\lambda = 1$ at $t = 0$, respectively Eq. (6) for $K = 6$ and Eq. (10) for $K = 8$ with $\eta^c$, is inserted to show the similarity and differences. Results of other Gr-systems are of similar fashion and not shown.

Fig. 2. The GrCP compacton branch requires rescaling to mitigate the issue of vanishing denominators, yet its outcomes are similar to those of the peakon branch and hence are not presented here. The "universality" is also in the sense that the major features from such a low $K$ (= 4) extend to large $K$s (see below), which is further demonstrated by the GrRH case with $K = 32$ where a sharp solitonic 'dark' (negative-sign) longon emerges subsequent to the apparent breakdown of the travelling-wave $u^\#$.

The whiskered torus, thus the "longon"carrying the interaction potential ${}^K\mathcal{G}$, echoes the image of "Long" (in Chinese Pinyin for the oriental dragon), the force carrier for various interactions between, say, human and nature, which embodies the transformation between finitude and infinity, upholding order amidst (potential) chaos, and so on and so forth. We have seen that such tori attract the heteroclinic orbits from nearby approximations starting from which it is possible to establish the a-posteriori KAM theorem [9], with only minor detailed differences in the proofs for different Gr-systems. We will henceforth focus on the GrBH case for further analysis.

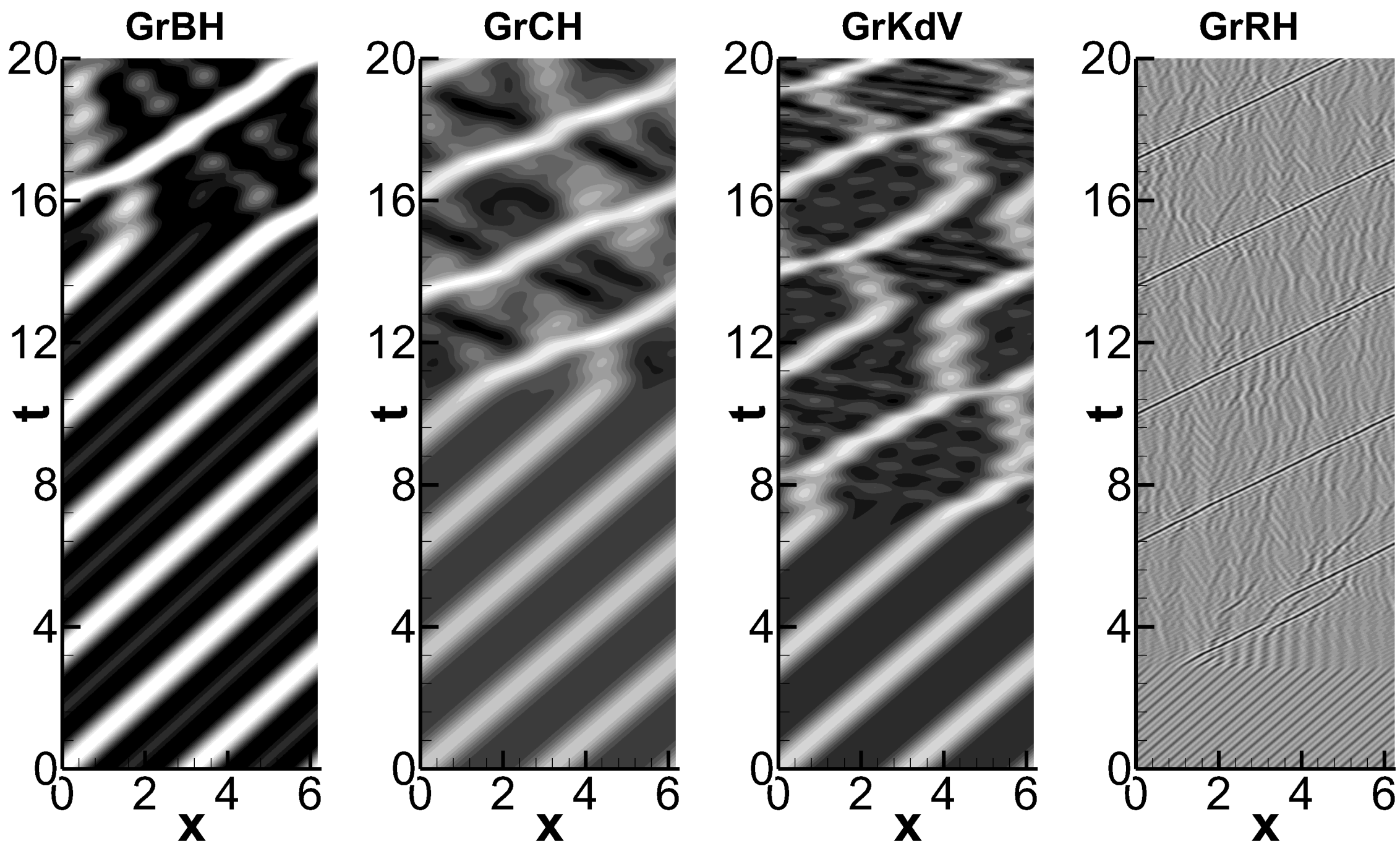

Figure 2. Space-time $u$-contours of the Gr-systems ($\lambda = 1$): GrBH with $K = 4$, GrCH with $K = 4$ (larger $u$ is brighter as can be read from Fig. 1's top frame; similarly for others), GrKdV with $K = 4$ and $\mu = -0.2$, and, GrRH with $K = 32$ and $\nu = -4$.

Gr-continuum on the lattice — Let the periodic lattice coordinate satisfy $x_j = x_{j+N}$, whence $v(x_{j+N}) = v(x_j) =: v_j$ for $j = 0,\ 1,\ 2,\ \ldots,\ N-1$, defining a discrete torus $\mathbb{T}_N$. The theoretical foundation of the standard (pseudo-)spectral method and the lattice representation of the GrBH continuum lies in replacing $\hat{v}_k$ defined earlier by the discrete Fourier transform (DFT) for $|k| \leq M$ (with $N-1 = 2M$ here), $\hat{\hat{v}}_k := \sum_{x_j \in \mathbb{T}_N} \frac{v_j}{N} e^{-\hat{i}kx_j} = \hat{v}_k + \sum_{i \neq 0} \hat{v}_{k+iN}$ (e.g., Ref. [15]). The aliasing error, represented by the second term, can be mitigated using dealiasing techniques like zero-padding or, alternatively speaking, truncation at $K < N/3$ ("2/3-rule").

Unifying the dealiasing and the Galerkin truncation results in, correspondingly, $\hat{\hat{u}}_k = \hat{u}_k$ for $u = P_K v$ in the GrBH equation (2), i.e., $\partial_t u_j = -P_K \partial_x u_j^2/2$, so

$$\partial_t u_j = \sum_{p+q=k\in{}^K\mathbb{G}}^{p,q\in{}^K\mathbb{G}} \sum_{x_m\in\mathbb{T}_N}^{x_n\in\mathbb{T}_N} \frac{k u_m u_n e^{\hat{i}(kx_j - px_m - qx_n)}}{2\hat{i}N^2} \tag{11}$$

where the right-hand side in physical-space variables reveals the GrBH lattice dynamics explicitly.

The pseudo-spectral method computes GrBH in Fourier space, evaluating the nonlinear term in physical space via DFT of $P_K u_j^2$, $\partial_t \hat{u}_k = -\frac{\hat{i}k}{2}\sum_{j=0}^{N-1}\frac{P_K(u_j^2)}{N}e^{-\hat{i}kx_j}$, so the computation aligns precisely with the GrBH definition, with only errors from the computer roundoff and time discretization. Note that even the "spectral-accuracy" errors, for the smooth solutions to classical differential equations (see also Ref. [16]), do not exist for us.

Since fourth-order Runge-Kutta scheme and its variant (for an approximation below) are used, the numerical results are highly accurate and reliable. Linear analysis (e.g., Lyapunov) of perturbed GrBH solutions (around, say, $u^\#$) indicates generic instability. And, conventional nonlinear analyses, such as orbital instability analysis, must contend with the unconventional ${}^K g$. Here, the numerical results potentially provide clues for further establishing relevant analytical insights.

Hamiltonian extrema and persistence — Notably, the critical set extremizing $\mathcal{H}$ via

$$\delta(\mathcal{H} - \lambda\mathcal{E})/\delta u = 0 \tag{12}$$

corresponds to Eq. (4). Including another multiplier $\lambda_0$ for $\mathcal{M}$, vanishing or not, doesn't matter, as in the KdV finite-gap theory [11, 12]. For the large-$K$ GrBH problem, there are in general ${}^K N_k = 2[K + \mathrm{sgn}(|\mathcal{M}|)] - 1 - |k|$ triads satisfying $p + q = k$ for each $k$. If ${}^K N_k$ were independent of $k$, then $\hat{u}_k = c(\lambda)$, a real constant uniform over $k$ would extremize $\mathcal{H}$. Therefore, for large $K$, with ${}^K N_k$ changing relatively slow with $k$,

$$u_0 = [\cos(x) + \cos(2x) + ... + \cos(Kx)]/\sqrt{K} \tag{13}$$

is an appropriate typical large-Hamiltonian but non-travelling-wave initial data for GrBH.

As illustrated in the left frame of Fig. 3 for the $u$-contour/carpet, the "universality" of the scenario of orbits attracted to the whiskered tori further extends to the above approximate $u_0$, and the sharper longon interactions with such high $K = 85$: two head-on colliding solitonic longons, with their strengths roughly proportional to the respective speeds, travel among the much weaker (c.f., middle-upper frame) and less-ordered ones; the latter present "long" quasi-trajectories with relatively severe "phase/position shifts" (a terminology borrowed from classical soliton theory for analogue) upon interaction (like the 'strange particles' and 'resonances' in particle physics): another reason for the term "longon". We may unify the emergence and decay of such 'particles' with the notion of chaotization and alternatively use terminologies such as 'thermalization', and any (statistically) stationary longulent state may also be termed '(statistical) equilibrium'. Viewing essentially all such oscillations as interacting solitonic longons may seem radical, yet it retains coherence and might found methematical justification for such "nonintegrable" systems (see more remarks on "pseudo-integrability" suggested below).

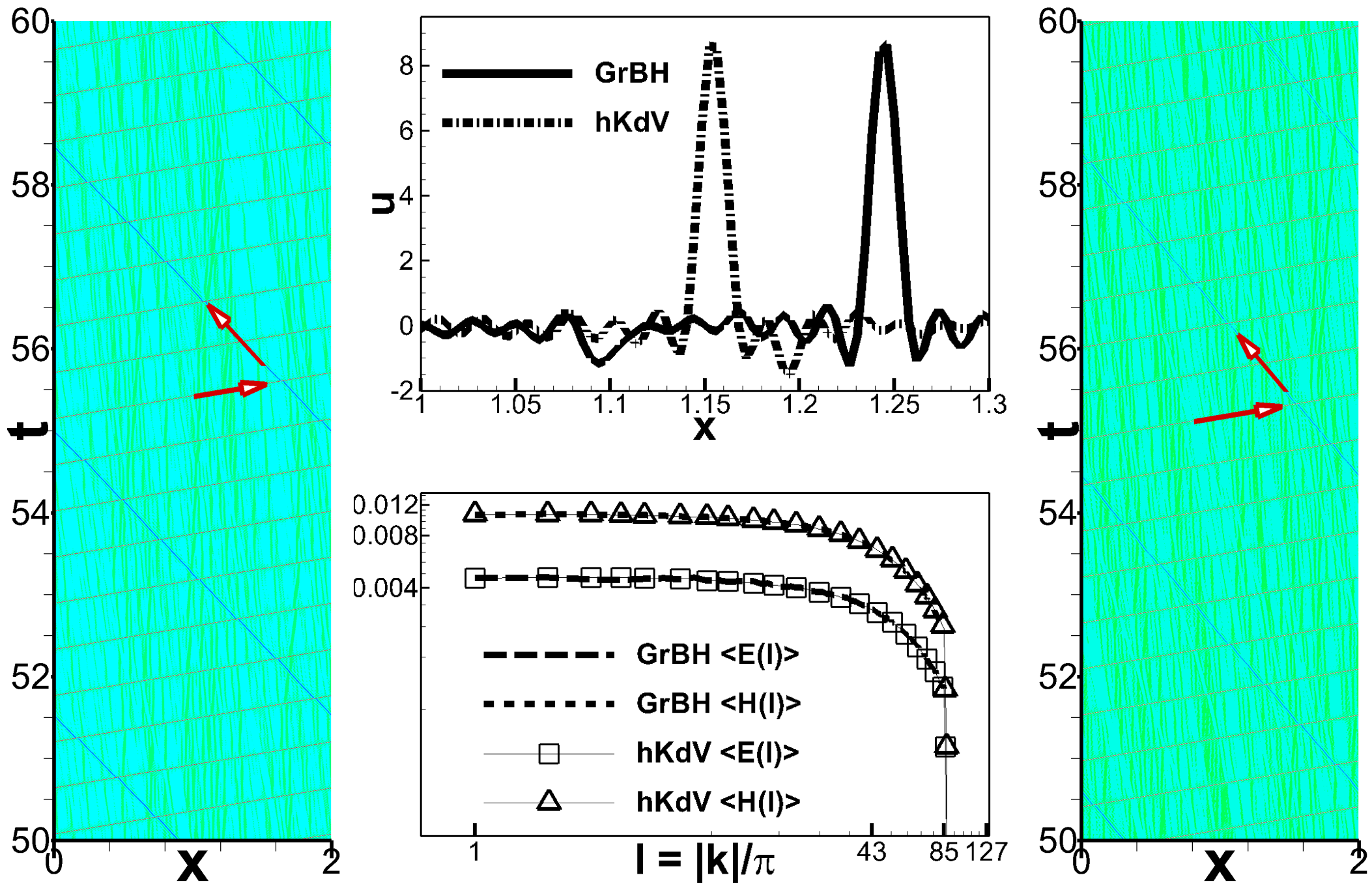

Figure 3. GrBH (left) and hKdV (right) $u$-contours. The period is normalized from $2\pi$ to 2. Snapshots (middle, upper) of the GrBH (at $t$ = 59.9) and hKdV (at $t$ = 59.7) $u$-profiles [corresponding to the contours whose color coding can be accordingly read], and, the energy and Hamiltonian spectra (middle, lower). The arrows are added to highlight the propagation of the apparently solitonic longons.

Besides the conventional energy spectrum $E(|k|) := \langle|\hat{u}_k|^2\rangle$, we may define

$$H(|k|) := \sum_p \langle \hat{u}_p \hat{u}_{k-p} \hat{u}_k^* + c.c.\rangle/6 \tag{14}$$

with $\langle\bullet\rangle$ for time averaging. The energy transfer rate is $T(|k|) := \hat{\imath} \sum_p \langle \hat{u}_p \hat{u}_{k-p} \hat{u}_k^* - c.c.\rangle/2$, showing some duality with $H(|k|)$. In GrBH absolute (statistical) equilibrium, $T = 0$ marks the balance of energy transfer, but $H$ provides additional information of the structures.

For an appropriate sequence $\omega_O$ of the "hyperdispersive"-KdV (hKdV) dispersive functions $\omega(n)$ of the model $\hat{a}_n = -\hat{\imath}\omega(n)\hat{v}_n$ in Eq. (1), the decoupled GrBH sub-dynamics with well-prepared $u_0$ in ${}^K\mathbb{G}$ can be approximated with $\omega_O(m) \to \infty$ for all $m \notin {}^K\mathbb{G}$ and $\omega_O(k) \to 0$ for all $k \in {}^K\mathbb{G}$. The asymptotic GrBH sub-dynamics may be argued directly by the fact that the intra- and extra-Galerkin frequencies can not match to form a resonant triad with a large jump of $\omega(n)$ in the classical resonant wave theory: the extra-Galerkin modes, if set up initially ("ill-prepared"), however, can have their own dynamics, not of the interest here though. For understanding some physics of dissipation, a choice of $a$ in Eq. (1) in Ref. [17] was the hyperviscous dissipation function $\propto -(k/k_G)^{2O}$ $(K < k_G < K+1)$ for integer $O \to \infty$, but more consistent with the current situation is the dispersive model such as the hKdV

$$\hat{a}_k = -\hat{\imath}\omega_O(k)\hat{v}_k;\ \ \omega_O = \begin{cases} (\frac{k}{k_G})^{2O+1} & \forall\ k \notin {}^K\mathbb{G} \\ 0 & \forall\ k \in {}^K\mathbb{G} \end{cases} \tag{15}$$

without the necessity of forcing for (energy) balance, among other subtleties. In the numerical computations [18] with correspondingly the same lattice number $N = 512$ and initial data, $k_G = 85.5 = K + 0.5$ and $O = 200$ are used in the hyper-dispersion Model (15). And, to avoid the slow change for $|k|$ near $k_G^+$, $\omega_O(k)$ is emperically set to be $750\,\mathrm{sgn}(k)(|k| - k_G)$ if $(|k|/k_G)^{401} < 1300$ (with the period normalized from $2\pi$ to 2 as in Fig. 3). We see in the middle-lower frame of Fig. 3 that $E$ and $H$, respectively, are close and show the equipartition tendency at small wavenumbers ($|k| < 10$, say): persistence of the very-high-dimensional whiskered tori associated to the longulent states are clearly indicated. Solitonic longon pulses approximate the Dirac delta function, thus the asymptotic large-scale energy equipartition; the nonlocal contribution to $H(|k|)$ at small $|k|$ from $p$ is dominated by small-$|p|$ modes, thus also equipartitioned $H(|k|)$.

Note $\hat{\imath}(k/k_G)^{401}\hat{u}_k$ corresponds to a Hamiltonian component $(\partial_x^{200}u)^2/(2k_G^{401})$ which however is minute, due to the smallness of $\hat{u}_m$ for all $m \notin {}^K\mathbb{G}$: the GrBH $\mathcal{H}$ is checked to be well preserved in the approximate model with tiny errors ($< 2\%$).

The hKdV structures shown in the right frame of Fig. 3 are also close to the GrBH ones (left frame). So, the results indicate the persistence, at least against hKdV(-like) models Eq. (15) with large-$O$ (with closer longulence for larger $O$ verified — not shown) as the perturbation to Gr, of the GrBH whiskered tori/longulent state, which is corroborated by other numerical results with different setups of various initial data, including the following zero-Hamiltonian ones.

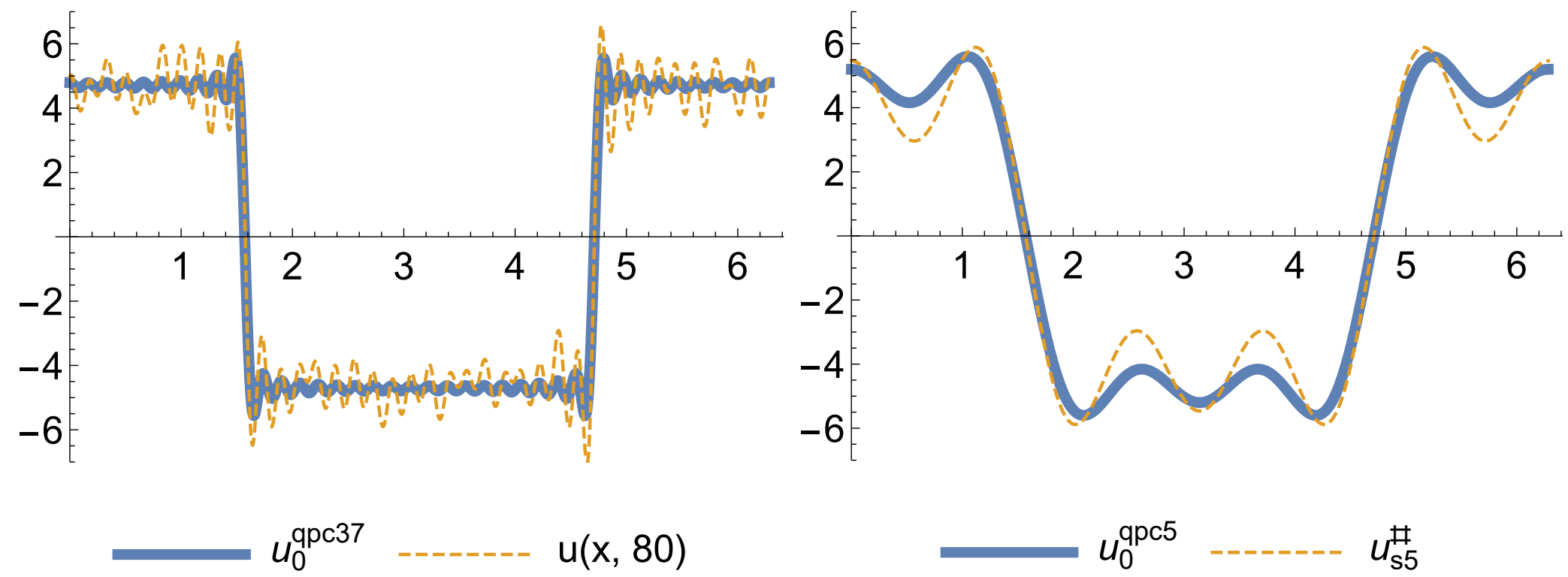

Figure 4. Well-developed $u(x,t)$ at a typical time $t = 80$ from $u_0^{qpc37}$ with $K = 37$ (left), and, $u_{s5}^{\#}$ and $u_0^{qpc5}$ (right); $Q = 3/2$ and $x_0 = -\pi/2$ for the QPC data.

Piecewise-constant $v_0 \sim \sum_k \frac{-2\hat{i}}{(2k+1)\pi} e^{\hat{i}(2k+1)x}$ is a weak solution to the BH equation, which suggests a quasi-piecewise-constant (QPC)

$$u_0^{qpcK} = Q \sum_{|2k+1| \le K} \frac{-2\hat{i}}{(2k+1)\pi} e^{\hat{i}(2k+1)(x+x_0)}, \tag{16}$$

parameterized with $Q$. Fig. 4 (left panel) shows, among other selectively thermalized or random-like weaker oscillations in the well-developed $u$ from $u_0^{qpc37}$, the persistent shock-antishock structure (as already in $u_0^{qpc37}$). The shock contributes a $E(|k|)$-component $\propto k^{-2}$ as already explicitly given in the zero-Hamiltonian $u_0^{qpcK}$ in Eq. (16).

On-torus invariants, quasi-periodic orbits, and a-posteriori KAM theorem — Further zero-$\lambda$ solutions to the solitary-wave equation (4) are considered for insights associated to the a-posteriori KAM theorem [9]: a stationary $u_{sK}^{\#}$ for $K = 37$ and slight deformation of it, responsible in the KAM fashion for the longulent $u(x,t)$ developed from $u_0^{qpc37}$ in Fig. 4, should be close by, just as the comparison between $u_0^{qpc5}$ and the stationary

$$u_{s5}^{\#} = 2(1 + \sqrt{3}) \cos x + 2 \cos(3x) + 2 \cos(5x), \tag{17}$$

calculated through the ansatz similar to Eq. (16) with $K = 5$. For large $K$, $u^{\#}_{sK}$s can be many, and it remains to identify the right one, including the possibility of other solutions with the replacement of the right-hand side of Eq. (4) by $\hat{i}\omega_K(k)\hat{u}^{\#}_k$. The longulence approximation (Fig. 3) with the linear dispersion (15), as a perturbation to ${}^K g$, indicates similar persistence mechanism working around.

The above KAM claim can be strengthened and made more explicit, with the formulation resembling that of classical integrable systems (such as KdV in, e.g., Refs. [11, 12]). The relevant tori carrying the quasi-periodic solutions "close" to $u^{\#}$ and the longulent solutions may be specified by a sub-set of invariants. With given invariants, we accordingly choose appropriate number of modes to be occupied with different frequencies, rationally-independent (incommensurate) or even Diophantine. For example, replacing the right-hand side of Eq. (4) by $-\hat{i}k(\lambda_0 + \lambda_1 k^2)\hat{u}^{\#}_k$, we can find various solutions which are quasi-periodic when $\lambda_0$ and $\lambda_1$ are rationally independent: this is realized by the variation principle

$$\delta(\mathcal{H} - \lambda_0\mathcal{E}_0 - \lambda_\tau\mathcal{E}_\tau)/\delta u = 0, \tag{18}$$

with $\mathcal{E}_0 = \mathcal{E}$ and with $\tau = 1$ in $\mathcal{E}_\tau = \int_0^{2\pi}(\partial_x^\tau u)^2/(4\pi)dx$: such $\mathcal{E}_\tau$ is invariant along the specified orbit (but, except for $\tau = 0$, in general should vary outside the torus), because of the invariance of $\mathcal{H}$ and $\mathcal{E}_0$.

Longulent patterns appearing closer to the exact solutions (compared to the one-phase travelling waves we have tested) can indeed be constructed with the help of addtional on-torus invariants, as shown in Fig. 5 (left panel) of the case with $K = 4$, starting from the data of frequency vector with $\lambda_0 = \sqrt(6)/10$ and $\lambda_1 = \pi/10$.

With $\mathfrak{T}$ invariants besides $\mathcal{H}$, including possibly other on-torus invariants than $\mathcal{E}_\tau$, a $\mathfrak{T}$-torus can be similarly realized. And, some torus specified as such may correspond to the longulent state such as that late-stage one in Fig. 3. That is, it might be expected that with the "right" additional on-torus invariant(s) a Gr-system becomes pseudo-integrable, in the sense of specifying precisely a longulent state.

Finally, remarks on the forcing-and-damping perturbation follow. As has been well-documented for various systems (including ordinary differential equations such as the Duffing equation), the forms of forcing and the matching of the (relative) strengths of forcing and damping determines different possibilities of the dynamical behaviors (with chaos, synchornization, invariant tori or fix point, involving concepts such as bifurcation). So, the construction of whiskered tori of Gr-systems with such perturbations is highly subtle, as briefly mentioned when Model (15) is introduced and indeed checked with numerical experiments. For the completeness purpose here for indicating the possibility of the robustness of the longulent states against appropriate forcing and damping, it suffices to showcase the example accompanying the above GrBH case associated to Eq. (18), leaving the more systematic analyses (including the computer-assisted proofs [16] of other results) for another communication.

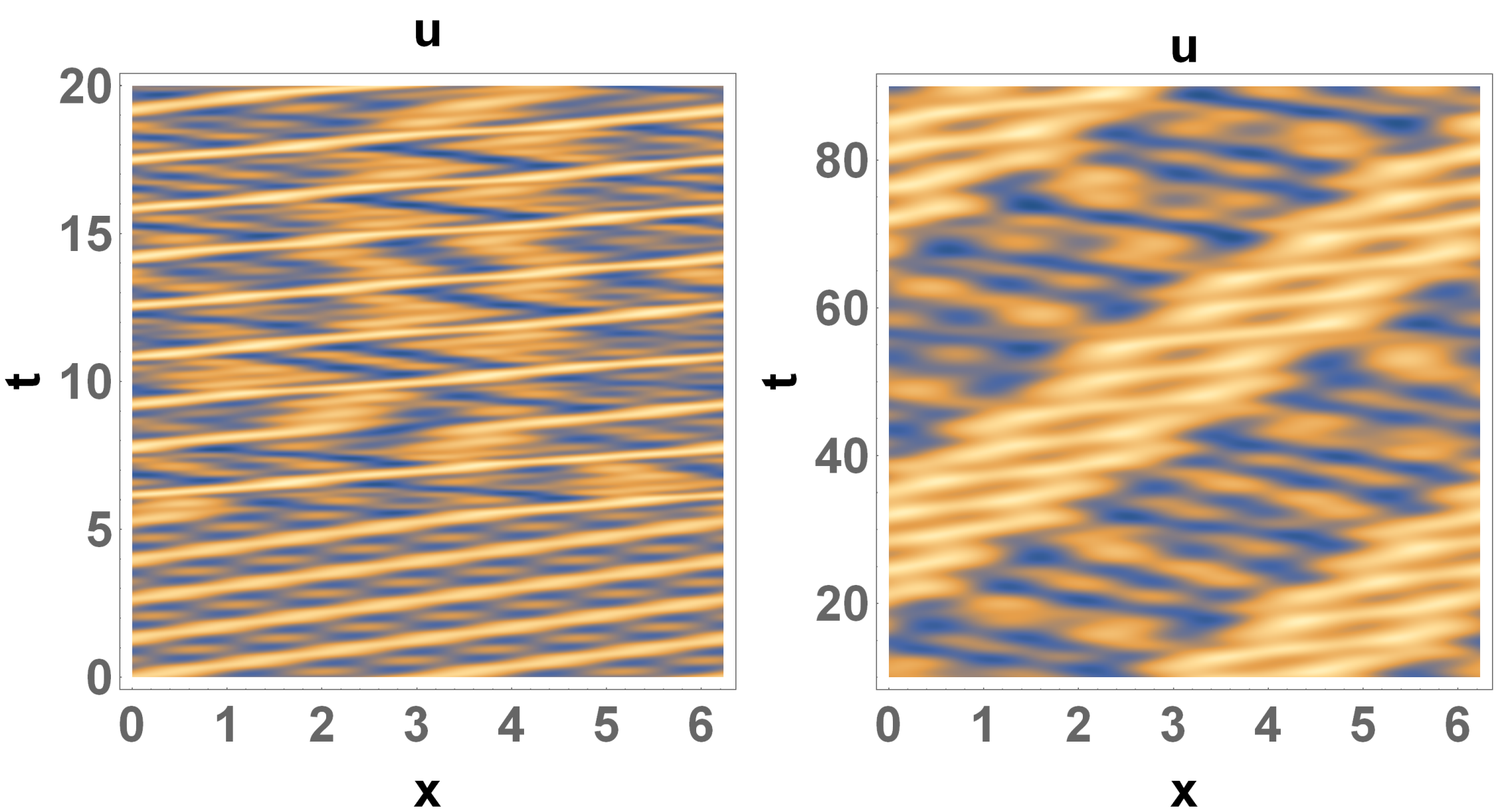

Figure 5. $u$-contours with two-frequency parameterization for the GrBH (left) and its forced-and-damped perturbation (right).

Fig. 5, right panel, presents the result from the same travelling-wave initial data which are also adopted to be the forcing up to a small coefficient. With the jump across $k_G$ being sharp and large enough, the details of the hyperviscous-type dissipation function, mentioned before Eq. (15), appear nonessential. Since both forcing and damping should be (very) small, it not well matched, it would need very long time to reach the final balanced state. We have thus manually adjusted the forcing and damping as time goes (ending with the forcing coefficient multiplying the initial data 0.003 and $O = 40$) to accelerate transition by, say, monitoring the variations of energy. But, still, it is not trivial to find appropriate parameters for producing patterns fairly close to the GrBH longulent state. Note that the whiskered tori may be supported by Cantorian parameter space (e.g., Ref. [20] for the complex Ginzburg-Landau —

CGL — case) the discreteness of which would add to the difficulty in numerical experiment. The forcing perturbation, to NLS, of CGL is "autonomous" for being depending on the dynamical variable, so the Cantorization may not carry over to the indepent forcing to GrBH here. We tend to believe that GrBH longulent states are persistent against appropriately matched damping and suitable forcing. The Galerkin regularization as an unusual pseudo-differential operator seems to make the rigorous proof more difficult. Nevertheless, the final pattern we show indicates the corresponding whiskered tori, in the fashion of unperturbed GrBH. And, rather than the mathematical details, the physical point we would like to point out and emphasize is that, just as the longons acting as a kind of intermittent structures leading to non-Gaussian GrBH statistics (and non-equipartition of the energy), particularly suggesting the dependence on the details of forcing of the intermittency of turbulence at Reynolds numbers however large but not infinite. The result is also reminiscent of the work of Chekhlov and Yakhot [21] who introduced the specific forcing to produce the Kolmogorov energy spectrum of Burgers turbulence, while we suggest persistent "longulence" here.

In conclusion, the author thanks J. M. Hyman for motivating communications.

* jz@sccfis.org